\documentstyle[prl,aps,amssymb,amsfonts,twocolumn,epsf]{revtex}
\def\1#1{{\bf #1}}
\def\2#1{{\cal #1}}
\def\4#1{{\tt #1}}
\def\5#1{{\sf #1}}
\def\6#1{{\frak #1}}
\def\7#1{{\Bbb #1}}
\def\8#1{{\rm #1}}
\def\9#1{{\cal #1}}

 \def\tr #1{{\mathop{\rm tr} \! \left( #1\right)}}
 \def\inn{_{\rm in}} \def\out{_{\rm out}}
  \def\idty{\11}
\title{\bf Semicausal operations are semilocalizable}
\author{T. Eggeling\thanks{Electronic Mail:
\tt{T.Eggeling@tu-bs.de}}, D. Schlingemann\thanks{Electronic Mail:
\tt{D.Schlingemann@tu-bs.de}} and R.F. Werner\thanks{Electronic Mail:
\tt{R.Werner@tu-bs.de}} \\
 {\small Institut f{\"u}r Mathematische Physik, TU Braunschweig,}\\
  {\small Mendelssohnstr.3, 38106 Braunschweig, Germany.}}

\begin{document}
\maketitle
\narrowtext

\begin{abstract} We prove a conjecture by DiVincenzo, which in the
terminology of Preskill et al.\ [quant-ph/0102043] states that
``semicausal operations are semilocalizable''.  That is, we show
that any operation on the combined system of Alice and Bob, which
does not allow Bob to send messages to Alice, can be represented
as an operation by Alice, transmitting  a quantum particle to Bob,
and a local operation by Bob.  The proof is based on the
uniqueness of the Stinespring representation for a completely
positive map. We sketch some of the problems in transferring these
concepts to the context of relativistic quantum field theory.
\end{abstract}

\section{Introduction} In a recent paper \cite{preskill} Preskill
et al. focus on the constraints that quantum operations must
fulfill in order to be compatible with relativistic quantum
theory. They introduce the notions of causal, localizable,
semicausal and semilocalizable quantum operations for bipartite
systems. The prefix {\em semi} refers to ``directed'' properties,
which are not invariant under exchanging Alice's and Bob's system.

\begin{figure}[h]
\begin{center}
\epsfxsize=5cm
\epsffile{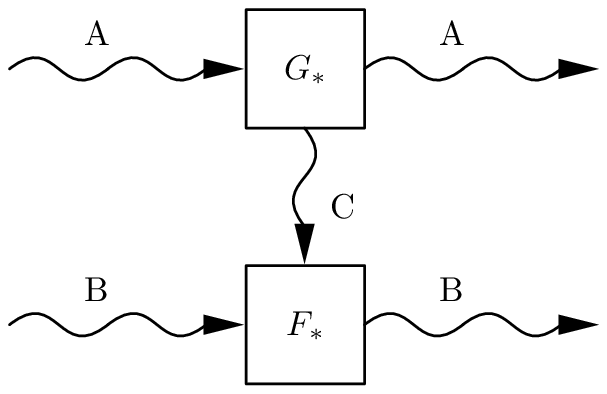}
\end{center}
\caption{}
\label{semilocalizable}
\end{figure}

An operation on a bipartite system is called  {\it
semilocalizable} for Alice, if it can be decomposed into two local
operations with one way quantum communication from Alice to Bob as
illustrated by Figure \ref{semilocalizable}: First Alice performs
a local operation $G$ on her system and sends quantum
information (via C) to Bob. Then Bob performs a local
operation $F$ on his system, depending on the information he got
from Alice.

An obvious consequence of this setup is that Bob cannot send
messages to Alice: the only possible carrier of information
(quantum or classical) is the system $C$, which goes from Alice to
Bob. Operations with this property will be called ``{\it
semicausal} ''. In other words, if we consider only measurements
on Alice's output of a semicausal operation, expectations do not
depend on Bob's initial preparation. That is, as far as Alice is
concerned, the device can be described by an operation on her
system alone.

The main result of this paper is the proof of the converse of the
above remark: if a device allows no signalling from Bob to Alice
(semicausality) we can represent it explicitly as a device
involving possibly a particle sent from Alice to Bob but none in
the other direction.

{\it Causality} and {\it localizablility} are defined as the
analogous symmetric properties. From this it would seem that these
are also equivalent (just use the proof twice, with an exchange of
the roles of Alice and Bob). However, full localizability is
defined to be stronger than the two semi-localizabilities: the
latter would mean only that there are two representations, each
involving only one-way communication, whereas localizability means
the absence of all communication. Indeed, in \cite{preskill} an
example for a causal operation which is not localizable has been
given.

Thus all proven implications between causal, semicausal,
localizable and semilocalizable operations (including the results
of this paper) can be visualized in the following diagram:
 \begin{equation}\label{arrows}
 \begin{array}{ccc}
{\rm localizable} & \Longrightarrow & {\rm causal}\\
     \Downarrow &        & \Downarrow\\
{\rm semilocalizable} & \Longleftrightarrow & {\rm
semicausal}
 \end{array} \end{equation}


As in \cite{preskill}, the concepts introduced so far only require
the standard setup of quantum information theory, in which
``localization'' is phrased entirely in terms of the Hilbert space
tensor product of Alice's and Bob's respective Hilbert space. This
kind of localization is a priori unrelated to relativistic
locality, and it turns out that for building a fully relativistic
quantum theory it is too narrow. Detailed knowledge about the
relativistic localization structures has been accumulated during
the last three decades in a research programme known variously as
``algebraic quantum field theory'' or ``local quantum physics''
\cite{araki,borchers,haag}. Quantum information theoretical
aspects have also been studied within this framework (See e.g.
\cite{werner,summerswerner,clifton} and references cited therein).
Technically, the main change is related to the fact that one has
to deal with infinitely many degrees of freedom, and occurs in a
similar way when discussing statistical mechanics in the
thermodynamic limit \cite{BraRo}: the observable algebra of a
local subsystem can no longer be taken as the operators of the
form $A\otimes\idty$ with respect to some tensor product
decomposition of the underlying Hilbert space. Instead, more
general von Neumann algebras must appear as observable algebras.
We will briefly comment on the changes this introduces in the
concepts of localized operations in the last section.

\section{Definition of localization properties}
\label{definitions}

As is well known, physical operations can be described either in
the Schr\"odinger picture, by a map acting on states or density
matrices or, equivalently,  in the Heisenberg picture, by an
operator acting on observables. In spite of this equivalence,
however, some things and especially localization properties and
the Stinespring dilation are stated more easily in the Heisenberg
picture. Therefore we will work in the Heisenberg picture.

We recall some basic notions and notations. In the Heisenberg
picture an {\it operation} \cite{davies} taking systems with
Hilbert space $\2H\inn$ to systems with Hilbert space $\2H\out$ is
a completely positive operator
$E\mathpunct:\6B(\2H\out)\to\6B(\2H\inn)$ satisfying
$E(\idty)\leq\idty$. The corresponding map $E_*$ in the
Schr\"odinger picture acts on a density matrix $\rho$ on $\2H\inn$
such that
\begin{equation}\label{duality}
   \tr{E_*(\rho)\;A}=\tr{\rho\; E(A)}
   \qquad \forall A\in\6B(\2H\out) \ .
\end{equation}
The conditions on $E$ are equivalent to  $E_*$ being likewise
completely positive, and satisfying the normalization condition
$\tr{E_*(\rho)}\leq1$ for all density operators $\rho$. If $E_*$
is even trace preserving (equivalently: $E(\idty)=\idty$), we call
the operation non-selective, or a {\it channel} \cite{fusz2}. As
the name suggests, selective operations typically occur, when part
of the operation is a selection of a sub-ensemble according to
measuring results obtained by the device.

The four properties in diagram~(\ref{arrows}) all refer to an
operation $E$ with
$\2H\inn=\2H\out=\2H_{AB}\equiv\2H_{A}\otimes\2H_{B}$.

\vspace{0.2cm}\noindent
\1{Definition:}
A completely positive map (not necessarily a channel)
$E\mathpunct:\6B(\2H_{AB})\to\6B(\2H_{AB})$ is called
\begin{enumerate}
\item {\em semicausal} if it can be written as
\begin{equation}\label{semicausal}
E(a\otimes \idty_B)=T(a)\otimes \idty_B
\end{equation}
for all $a\in\6B(\2H_A)$ and for some completely positive map
 $T\mathpunct:\6B(\2H_A)\to\6B(\2H_A)$.
\item {\em causal} if it semicausal in both directions, i.e. if
\begin{eqnarray}\label{causal}
E(a\otimes \idty_B)&=&T(a)\otimes \idty_B \qquad \mbox{and}\\
E(\idty_A\otimes b)&=&\idty_A\otimes T'(b)
\end{eqnarray}
for all $a\in\6B(\2H_A),$ $b\in\6B(\2H_B)$ and for some completely positive maps
$T\mathpunct:\6B(\2H_A)\to\6B(\2H_A)$ and $T'\mathpunct:\6B(\2H_B)\to\6B(\2H_B).$
\item {\em localizable} if it can be decomposed into
\begin{equation}\label{local}
E=G\otimes F
\end{equation}
where $F\mathpunct:\6B(\2H_{B})\to \6B(\2H_{B})$ is a channel
and $G\mathpunct:\6B(\2H_{A})\to\6B(\2H_A)$ a
completely positive map.
\item {\em semilocalizable} if it can be decomposed into
\begin{equation}\label{semilocal}
E=(G\otimes \8{id}_B)\circ(\8{id}_A\otimes F)
\end{equation}
where $F\mathpunct:\6B(\2H_{B})\to \6B(\2H_{CB})$ is a channel,
$G\mathpunct:\6B(\2H_{AC})\to\6B(\2H_A)$ a completely positive map and
$\2H_C$ a finite dimensional Hilbert space.
\end{enumerate}

\section{Semicausal operations are  semilocal} \label{mainpart}

We are now prepared to give the proof of the conjecture by
DiVincenzo, of which the special case of complete measurements was
treated in \cite{preskill}.

\vspace{0.2cm}\noindent {\bf Theorem:} {\em A completely positive
map (not necessarily a channel) is semilocal if and only if it is
semicausal.} \vspace{0.2cm}

Before going into the proof, we point out some salient facts about
the Stinespring representation of a completely positive map
\cite{stinespring}.

 \paragraph{The Stinespring representation:} The Stinespring representation
theorem, as adapted to maps between finite dimensional quantum
systems, states that any completely positive map
$E\mathpunct:\6B(\2H\out)\to\6B(\2H\inn)$  can
be written as
 \begin{equation}\label{stein1}
        E(a)=V^*(a\otimes\idty_\2K)V
 \end{equation}
 with a linear operator
$V\mathpunct:\2H\inn\to\2H\out\otimes\2K$, where $\2K$ is a finite
dimensional Hilbert space, called the dilation space. The
representation (\ref{stein1}) is called minimal if (and only if)
the set of vectors
 \begin{equation}
      (a\otimes\idty_\2K)V\varphi
 \end{equation}
with $a\in\6B(\2H\out)$ and $\varphi\in\2H\inn$ spans
$\2H\out\otimes\2K$.

\paragraph{Uniqueness:} The main step of the proof will be to get
a factorization of the given operation into two operations with
suitable localization properties. It turns out that such a
factorization is provided precisely by the uniqueness statement
for the Stinespring dilation. We therefore explain this uniqueness
in more detail.

Suppose that $E$ has a minimal Stinespring representation
(\ref{stein1}) as well as a further (not necessarily minimal) one
 \begin{equation}\label{stein2}
        E(a)=V_1^*(a\otimes\idty_{\2K_1})V^{}_1
 \end{equation} with another
linear map $V_1\mathpunct:\2H\to\2H\otimes\2K_1$. Since
representation (\ref{stein1}) is taken to be minimal, we conclude
$\8{dim}(\2K)\leq \8{dim}(\2K_1)$ and the prescription
 \begin{equation}\label{dilavec} \tilde
   U(a\otimes\idty_{\2K})V\psi:= (a\otimes\idty_{\2K_1})V_1\psi
 \end{equation}
  yields a well defined isometry
$\tilde U\mathpunct:\2H\otimes\2K\to\2H\otimes\2K_1$. This can be
easily verified by observing that all scalar products between
vectors such as the ones on the right hand side of (\ref{dilavec})
are fixed by the relation (\ref{stein2}). From this definition of
$\tilde U$ we find that the intertwining relation
 \begin{equation}
   \tilde U(a\otimes\idty_{\2K})=(a\otimes\idty_{\2K_1})\tilde U
 \end{equation}
 holds for each $a\in\6B(\2H)$. Hence $\tilde U$
must be decomposable into
 \begin{equation}\label{iso}
     \tilde U=\idty_\2H\otimes U
 \end{equation} with an isometry
$U\mathpunct:\2K\to\2K_1$. If
both representations (\ref{stein1}) and (\ref{stein2}) are minimal
the dimensions of the dilation spaces coincide and $U$ must be a
unitary operator. The minimal Stinespring representation is thus
unique up to a unitary transformation.

\paragraph{Kraus operators:} In most of the current literature the
Stinespring dilation is used only in the form of a corollary,
called the Kraus representation of a completely positive map. If
we introduce an  orthonormal basis
 $(\varepsilon_1,\cdots, \varepsilon_k)$ of the dilation space
$\2K$ and define the ``Kraus operators''
$K_\alpha\mathpunct:\2H\inn\to\2H\out$ by
 \begin{equation}
 V\psi=\sum_\alpha (K_\alpha\psi)\otimes\varepsilon_\alpha \quad
         \forall\psi\in\2H\inn,
 \end{equation}
we can write the dilation formula~(\ref{stein1}) as
 $E(a)=\sum_{\alpha=1}^k K_\alpha^* a K^{}_\alpha$
 for all $a \in\6B(\2H\out)$.
Of course, everything we do in this paper could be formulated in
terms of Kraus operators. We found this less practical, however,
because the above uniqueness statement becomes more involved: The
choice of the basis $\varepsilon_\alpha$ always introduces some
arbitrariness, so even in the minimal case the collection of Kraus
operators is only unique up to a unitary transformation acting on
the index $\alpha$.

\vskip20pt
 \noindent\1{Proof of Theorem:} As noted in the introduction
the implication ``semilocalizable$\Longrightarrow$semilocal'' is
trivial. In the notation of Section~\ref{definitions} it becomes
$$E(a\otimes \idty_B)=G(a\otimes\idty_C)\otimes \idty_B$$ so
(\ref{semicausal}) follows with $T(a):=G(a\otimes\idty_C)$.

For the reverse implication, let
$E\mathpunct:\6B(\2H_{AB})\to\6B(\2H_{AB})$ be a semicausal
operation, i.e. an operation fulfilling (\ref{semicausal}) for
some completely positive map
$T\mathpunct:\6B(\2H_A)\to\6B(\2H_A)$. The Stinespring
representation of $E$ gives us a dilation space $\2H_D$ and a
linear operator $V\mathpunct:\2H_{AB}\to\2H_{ABD}$,
$\2H_{ABD}\equiv\2H_{AB}\otimes\2H_D$, such that
\begin{equation}\label{stein} E(a\otimes b)=V^* (a\otimes b\otimes
\idty_D)V \end{equation} for each $a\in\6B(\2H_A)$ and
$b\in\6B(\2H_B)$. Analogously we get a dilation space $\2H_C$ and
a linear operator $W\mathpunct:\2H_A\to\2H_{AC}$ from the minimal
Stinespring representation of $T:$
 \begin{equation}
    T(a)=W^*(a\otimes\idty_C)W
 \end{equation} for all $a\in\6B(\2H_A)$.
According to the relation (\ref{semicausal}) we obtain
\begin{equation}
    V^*(a\otimes\idty_{BD})V
      = (W^*\otimes \idty_B)(a\otimes\idty_{CB})(W\otimes\idty_B)
 \end{equation} for
all $a\in \6B(\2H_A)$ \cite{fusz5}. The uniqueness of the minimal
Stinespring representation now implies the existence of an
isometry $U\mathpunct:\2H_{CB}\to\2H_{BD}$ (see Eq.~(\ref{iso}))
such that \begin{equation} (a\otimes\idty_{BD})V= (\idty_A\otimes
U)(a\otimes\idty_{CB})(W\otimes\idty_B) \end{equation} and
therefore
 \begin{equation}\label{compo}
   V=(\idty_A\otimes U)(W\otimes\idty_B) \; .
 \end{equation}
 From this we obtain a completely positive unital map
$F\mathpunct:\6B(\2H_B)\to\6B(\2H_{CB})$ by taking
\begin{equation}
 F(b):=U^*(b\otimes\idty_D)U
\end{equation} for
every $b\in\6B(\2H_B)$ and a completely positive (not necessarily
unital) map $G\mathpunct:\6B(\2H_{AC})\to\6B(\2H_{A})$:
\begin{equation}
   G(a\otimes c):=W^*(a\otimes c)W
\end{equation}
for every $a\in\6B(\2H_A)$ and $c\in\6B(\2H_C)$. Thus we conclude
from (\ref{stein}) and (\ref{compo}) that the identity
(\ref{semilocal}) holds. In other words, $E$ is
semilocalizable. \hfill$\Box$

\section{Outlook} \label{conclusions}

As already indicated in the introduction, quantum field theory
requires a more general setup than the one used in this note and
in \cite{preskill}. Relativistic localization is then expressed by
assigning to every spacetime region the algebra of observables,
which can be measured in that region, typically a von Neumann
algebra. Signal causality then means that whenever two regions are
spacelike separated (no causal signals can be exchanged) the
corresponding algebras commute elementwise. One might try
replacing this  by an assignment of ``local Hilbert spaces'' to
spacetime regions, such that the union of spacelike separated
regions corresponds to the Hilbert space tensor product. This is
not possible, however, because spacetime regions can be split into
finer and finer pieces, and this would create difficulties for the
tensor products, especially when the overall Hilbert space is
required to be separable (to have a countable orthonormal basis),
and relativistic invariance is imposed.

Surprisingly the von Neumann algebras of local regions are all
isomorphic under mild axiomatic assumptions. More specifically
they are all isomorphic to the unique hyperfinite type
III$_1$-factor (see \cite{baumwoll} for an explanation of these
terms, and the proof). Nevertheless the localization structure is
far from trivial, and resides in the way these algebras are nested
into each other. Already for the inclusion of two such algebras an
amazing variety is possible.

As might be expected from the heuristic argument at the beginning
of this section, the small distance localization structure in
quantum field theory deviates dramatically from what would be
expected from Hilbert space tensor products. Suppose Alice and Bob
operate in spacelike separated regions, and $\6A$ and $\6B$ are
the local von Neumann algebras assigned to these regions. Then
when the regions are very close together, no physical
\cite{fuszNormal} product states exist, hence there are no
separable states at all. In fact, all physical states violate the
CHSH-Bell inequality maximally \cite{summerswerner}.

On the other hand, if the regions are a finite distance apart, the
so-called {\it split property} holds, which is equivalent to the
existence of many separable states, or to the possibility for Bob
to prepare any state of his subsystem locally without disturbing
Alice's \cite{werner}. Algebraically this means that the von
Neumann algebra generated by $\6A$ and $\6B$ is isomorphic to the
von Neumann algebra tensor product $\6A\overline\otimes\6B$
\cite{sakai}.

The notion of {\it semicausality} is easy to express in this
context: that $E$ maps operators of the form $a\otimes\idty$ to
operators of the same form just means that Alice's algebra is
invariant in the sense that $a\in\6A \Longrightarrow E(a)\in\6A$,
or briefly $E(\6A)\subset\6A$.

It is not so clear what should be understood by {\it
semilocality}. The reason is that one has to decide what degree of
independence should be postulated for the intermediate system $C$,
sent from Alice to Bob in Figure~\ref{semilocalizable}. If we
require the kind of independence valid for widely separated
regions (split property), as suggested by the image of a system
being ``sent'', the Theorem is probably false. Although the
Stinespring decomposition for semicausal maps is well understood,
it is not clear what factorizations would be meaningful
interpretations of Figure~\ref{semilocalizable}.

\section*{Acknowledgements} Funding by the European Union project
EQUIP (contract IST-1999-11053) and financial support from the DFG
(Bonn) are gratefully acknowledged.


\begin{references} \bibitem{preskill} D. Beckman, D. Gottesman, M.
A. Nielsen and John Preskill,
 quant-ph/0102043 (2001).

\bibitem{araki} H. Araki, Mathematical theory of quantum fields,
Oxford University Press, (1999)

\bibitem{borchers} H.-J. Borchers, Translation group and particle
representations in quantum field theory,
 Lecture Notes in Physics (Springer Verlag, Berlin), 1996.

\bibitem{haag} R. Haag, Local quantum physics, Springer-Verlag,
Berlin 1992. (A second edition was released in 1996.)

\bibitem{werner} R.F. Werner, Local preparability of states and
the split property in quantum field theory,  Lett. Math. Phys.
{\bf 13} (1987) 325-329.

\bibitem{summerswerner} S.J. Summers, R.F. Werner, Maximal violation
of Bell's inequalities for algebras of observables in
tangent spacetime regions, Ann. Inst. H. Poincar\'e, {\bf A 49} (1988) 215-243.

\bibitem{clifton} R. Clifton, H. Halvorson, Entanglement and open systems in
algebraic quantum field theory,
quant-ph/0001107.

\bibitem{BraRo} O. Bratteli and D.W. Robinson,
Operator algebras and quantum statistical mechanics, two volumes,
(Springer Verlag, New York) 1979 and 1981.

\bibitem{davies} E. B. Davies, Quantum Theory of Open Systems,
Academic Press 1976.

\bibitem{fusz2} In \cite{preskill} the term {\it superoperator} is
used for (the dual map of) a {\em channel}.

\bibitem{stinespring} W. F. Stinespring, Proc. Amer. Math. Soc.
{\bf 6}, 211-216 (1955).

\bibitem{fusz5} Note that if $W$ is a minimal dilation of $T$ then
so is $W\otimes\idty_B$ for E.

\bibitem{fuszNormal}Here ``physical state'' technically means a
``normal'' state, i.e., a state which can be represented by a
density operator on the Hilbert space of the global system, as
opposed to ``singular'' states. For the purposes of this
discussion it is sufficient to require ``local normality'', i.e.,
that the restriction to bounded spacetime regions can be
represented in this way. Roughly speaking this requires that no
infinite amount of energy needs to be invested in a finite region
to prepare the state. This property allows also temperature states
of finite density, which are globally different (singular) from
the vacuum, but not locally.

\bibitem{baumwoll} H. Baumg\"artel and M. Wollenberg, Causal nets
of operator algebras, Akademie Verlag Berlin, 1992.

\bibitem{sakai} S. Sakai, C*-algebras and W*-algebras, Classics in
Mathematics, Springer-Verlag Berlin Heidelberg New York, 1971.


\end{references}
\end{document}